\documentclass [11pt]{article}
\usepackage{setspace,amsfonts,epsfig,amssymb,mathrsfs,amsmath}
\usepackage[margin=1.0in]{geometry}
\usepackage{hyperref}
\usepackage{color}

\pdfoutput=1

\begin{document}

\title{In search of lost spacetime: philosophical issues arising in quantum gravity}
\author{Christian W\"uthrich}
\date{To appear in Soazig Le Bihan (ed.), {\em La Philosophie de la Physique: D'aujourd'hui \`a demain}, Paris : Vuibert.}
\maketitle

\begin{abstract}\noindent
This essay presents an accessible introduction to the basic motivations to seek a quantum theory of gravity. It focuses on one approach---loop quantum gravity---as an example of the rich philosophical issues that arise when we try to combine spacetime and quantum physics.
\end{abstract}

\noindent
Theoretical physics today is confronted with a challenge reminiscent of the one Newton's predecessors faced in the 17th century: two incompatible theories quite successfully describe two different domains of phenomena. The laws of quantum mechanics govern the small-scale phenomena of elementary particle physics, and the laws of general relativity (GR) encode the large-scale structure of the universe. The present challenge of quantum gravity is that of completing the revolution that took place in physics in the last century; the task is no less than to fuse the two incommensurable frameworks of quantum physics and GR. Many protagonists in this endeavour hope that meeting this challenge will amount to a substantive, and perhaps final, step toward the theoretical unification in fundamental physics. This Herculean task has attracted more physicists today than ever before, plugging the ground, prospecting to find the gold mine leading to the holy grail. Their efforts have yielded a rich variety of approaches, techniques, and theories and include, most prominently, string theory and loop quantum gravity (LQG). Despite these exciting new developments in physics, philosophers have been remarkably slow at engaging the conceptually and philosophically rich material that has been unearthed in the process. 

This paper issues a call to arms and seeks to entice the reader with some of the most captivating philosophical puzzles arising in quantum gravity. The analysis will be prefaced, in Section \ref{sec:why}, by general considerations concerning the need for finding a quantum theory of gravity and the methods used in the pursuit of this goal. After mapping the field in Section \ref{sec:mapping}, I will introduce LQG as an important competitor and particularly rich source of philosophical trouble in Section \ref{sec:lqg}. The so-called problem of time, i.e.\ the puzzle that no genuine physical quantity can change, is discussed in Section \ref{sec:time}. Finally, Section \ref{sec:emerge} analyzes how the familiar continuous spacetime structure might re-emerge from the fundamental, non-spatiotemporal structure.

\section{Why quantum gravity?}
\label{sec:why}

Before we embark upon an investigation of the foundations of quantum gravity, we ought to convince ourselves that a theory of quantum gravity is indeed necessary. A {\em quantum theory of gravity}  is any consistent theory which combines gravity with a quantum description of matter. It is important to note that this does not entail that a quantum theory of gravity must regard gravity itself as quantized. It is at least conceivable that such a theory marries a classical understanding of gravity with a quantum understanding of matter. As important methodological constraints on a quantum theory of gravity, we demand that it has the ``appropriate limits,'' i.e.\ at scales where the quantum nature of matter becomes irrelevant, the theory ought to merge into GR, and in regimes in which gravity is weak, it ought to turn into a quantum theory.\footnote{More precisely, it ought to turn into a quantum field theory on the Minkowski spacetime of special relativity.} 

But why does physics need a quantum theory of gravity at all? The usual answer to this question is a combination of the following three (groups of) arguments. First, it is often claimed that such a theory is necessitated by a demand for {\em unification}. Based on its successful history in physics, unification has exerted a great methodological attraction to many. James Clerk Maxwell unified electric and magnetic forces into a dynamical theory of electromagnetism in the 1870s. In the 1960s, Abdus Salam, Sheldon Glashow, and Steven Weinberg formulated the electroweak theory, unifying electrodynamics and the weak nuclear force associated with radioactivity. Next, quantum chromodynamics, describing the strong nuclear force binding the nuclei of atoms and their constituents, and the electroweak theory have been unified into the standard model of particle physics, which successfully accounts for three of the four fundamental forces in physics. It is thus a natural ambition to attempt a unification of the quantum theories of the standard model with GR, our currently best theory of the remaining force---gravity. While its venerable history legitimizes unification as a methodological desideratum---and, to some extent, as a research programme---, it does not justify it as a metaphysical dogma. The past success of unification does not entail that nature must be sufficiently unified as to allow a single fundamental theory to underwrite all of physics. Thus, it remains perfectly conceivable that nature is disunified in the sense that gravity is not subsumable under the quantum umbrella of particle physics. 

The second answer trades on the singularity theorems proven in the 1960s and 1970s by Stephen Hawking, Roger Penrose, and Robert Geroch, which firmly establish that singularities are generic in classical GR. Many authors have argued that GR loses its validity ``there'' and that it thus contains the seeds of its own destruction. Therefore, the argument goes, a replacement is needed and a quantum theory of gravity is expected to fill that gap. More particularly, quantizing gravity, i.e.\ describing gravity as having a quantum nature itself is believed by some to dissolve singularities such as the big bang. But why should this argument, at least by itself, have much force? In GR, singularities are not part of the spacetime fabric, i.e.\ they are not ``at'' a particular ``location,'' and hence there is no need to have a valid theory ``there'' as far as GR is concerned. GR is a perfectly consistent theory within the realm of its applicability and it does not, therefore, contain the seeds of its own destruction. At least, these seeds cannot bear any dialectical fruit without a whole lot of additional argumentative fertilizer. 

Third, and by far most compellingly, there are phenomena for which we have good reason to believe that both gravity as well as quantum effects matter and that thus both are ineliminable ingredients to a theory which successfully describes these phenomena. Most importantly, these phenomena include the dynamics of black holes and the very early universe. It is important to appreciate that while both phenomena involve---in their classical description---a singularity, the necessity for a quantum theory of gravity does not arise because of this. Rather, the small scales and the high densities of matter and the simultaneous presence of a strong gravitational field jointly necessitate such a theory. Ultimately, it is thus the existence of rather extreme phenomena, and not some methodological or aesthetic criteria, that drive the need for a quantum theory of gravity. 

Note that while quantizing gravity---if successful---lends itself rather straightforwardly to a quantum theory of gravity, it is not necessary to obtain a quantum theory of gravity. The existence of regimes in which both quantum effects of matter and strong gravitational fields play an important role does not imply that gravity must be quantum itself. Instead, we only need a theory that governs the ``interaction'' between the quantum matter and the possibly classical gravity. In other words, so-called ``semi-classical'' theories of gravity have not been ruled out by anything said up to this point, even though they violate principles of GR. 

Having driven the wedge between the issues of whether we need a quantum theory of gravity and whether gravity needs to be quantized, I hasten to add that there are a number of arguments that pertain to show that in any quantum theory of gravity, gravity must be quantized and that thus, semi-classical approaches are not feasible. Typically, these arguments attempt to derive a contradiction with a well-entrenched physical principle such as the correspondence principle or the conservation of energy from the assumption of classical gravity interacting with quantum matter. However, I am not aware of any such argument which fully succeeds without invoking additional assumptions that an advocate of a semi-classical approach need not accept.\footnote{Cf.\ Callender and Huggett (2001), Huggett and Callender (2001), Mattingly (2006), and W\"uthrich (2005).}

Having secured the need for a quantum theory of gravity, let me then press on to briefly map the main competing approaches to quantum gravity.

\section{Mapping the field: approaches to quantum gravity}
\label{sec:mapping}

Introducing a helpful taxonomic scheme, Chris Isham (1994) proposed to divide the many approaches to formulating a full, i.e.\ not semi-classical, quantum theory of gravity into four broad types of approaches: first, those quantizing GR; second, those ``general-relativizing'' quantum physics; third, construct a conventional quantum theory including gravity and regard GR as its low-energy limit; and fourth, consider both GR and conventional quantum theories of matter as low-energy limits of a radically novel fundamental theory. Let us briefly consider each group in turn.

The first family of strategies starts out from classical GR and seek to apply, in a mathematically rigorous and physically principled way, a ``quantization'' procedure, i.e.\ a recipe for cooking up a quantum theory from a classical theory such as GR. Of course, quantization proceeds, metaphysically speaking, backwards in that it starts out from the dubious classical theory---which is found to be deficient and hence in need of replacement---and tries to erect the sound building of a quantum theory of gravity on its ruin. But it should be understood, just like Wittgenstein's ladder, as a methodologically promising means to an end. Quantization procedures have successfully been applied elsewhere in physics and produced, among others, important theories such as quantum electrodynamics. Advocates of approaches in this family hope to repeat these successes in gravitational physics. 

The first family consists of two genera, the now mostly defunct covariant ansatz\footnote{Defunct because covariant quantizations of GR are not perturbatively renormalizable, a flaw usually considered fatal. This is {\em not} to say, however, that covariant techniques don't play a role in contemporary quantum gravity.} and the vigorous canonical quantization approach. A canonical quantization requires that the theory to be quantized is expressed in a particular formalism, the so-called constrained Hamiltonian formalism. How casting GR as a constrained Hamiltonian system lies at the heart of its most perplexing conceptual issues will be discussed below. Loop quantum gravity (LQG) is the most prominent representative of this camp, but there are other approaches.

Secondly, there is to date no promising avenue to gaining a {\em full} quantum theory of gravity by ``general-relativizing'' quantum (field) theories, i.e.\ by employing techniques that permit the full incorporation of the lessons of GR into a quantum theory. The only existing representative of this approach consists of attempts to formulate a quantum field theory on a curved rather than the usual flat background spacetime. The general idea of this approach is to incorporate, in some local sense, GR's principle of general covariance. It is important to note that, however, that the background spacetime, curved though it may be, is in no way dynamic. In other words, it cannot be interpreted, as it can in GR, to interact with the matter fields.

The third group also takes quantum physics as its vantage point, but instead of directly incorporating the lessons of GR, attempts to extend quantum physics with means as conventional as possible in order to include gravity. GR, it is hoped, will then drop out of the resulting theory in its low-energy limit. By far the most promising member of this family is string theory, which, however, goes well beyond conventional quantum field theory, both methodologically and in terms of ambition. Despite its extending the assumed boundaries of the family, string theory still takes conventional quantum field theory as its vantage point, both historically and systematically, and does not attempt to build a novel theory of quantum gravity dissociated from ``old'' physics. Again, there are other approaches in this family, such as topological quantum field theory, but none of them musters substantial support among physicists.

The fourth and final group of the Ishamian taxonomy is most aptly characterized by its iconoclastic attitude. For the heterodox approaches of this type, no known physics serves as starting point; rather, radically novel perspectives are considered in an attempt to formulate a quantum theory of gravity {\em ab initio}. As far as I aware, these approaches only suggest programmatic {\em schemes}, rather than full-fledged theories. They derive their attraction mostly from the daunting appearance of deep incompatibility of the guiding principles of the quantum physics of the very small and the GR of the very large. This incompatibility, it is argued, cannot be resolved unless a radically fresh start is undertaken.

All these approaches have their attractions and hence their following. But all of them also have their deficiencies. To list them comprehensively would go well beyond the present endeavour. Apart from the two major challenges for LQG, which I will discuss subsequently, I shall be content to emphasize that a major problem common to all of them is their complete lack of a real connection to observations or experiments. There are some proposals how some or all approaches may make contact with the empirical, but so far these proposals do not go beyond often rather speculative suggestions of how such contact may be established. Either the theory is too flexible so as to be able to accommodate almost any empirical data, such as string theory's predictions of supersymmetric particles which have been constantly revised in light of particle detectors' failures to find them at the predicted energies or as string theory's {\em embarras de richesses}, the now notorious ``landscape problem'' of choosing among $10^{500}$ different models. Or the connection between the mostly understood data and the theories is highly tenuous and controversial, such as the issue of how---and whether---data narrowly confining possible violations of Lorentz symmetry relate to theories of quantum gravity predicting or assuming a discrete spacetime structure that is believed to violate, or at least modify, the Lorentz symmetry so well confirmed at larger scales. Or the predictions made by the theories are only testable in experimental regimes so far removed from present technological capacities, such as the predictions of LQG that spacetime is discrete at the Planck level at a quintillion ($10^{18}$) times the energy scales probed by the Large Hadron Collider at CERN. Or simply no one remotely has a clue as to how the theory might connect to the empirical, such as is the case for the inchoate approaches of the fourth group like causal set theory.

\section{Introducing loop quantum gravity}
\label{sec:lqg}

LQG is, apart from string theory, the most important approach to quantum gravity, both in terms of promise and of numbers of followers. It is a canonical approach which takes classical GR---our best classical theory of gravity---as its starting point and applies a well-tested procedure of cooking up a quantum theory from a classical theory in the hope that this will result in a viable quantum theory of gravity. It is an essentially conservative approach in that it aspires to remain as faithful as possible to known and successful physics. 

For the quantization procedure of choice---the so-called {\em canonical quantization}---to get some traction, it is necessary to reformulate GR as a Hamiltonian system. A Hamiltonian system is a physical system that obeys Hamilton's equations, which are the differential equation relating the (generalized) positions and the (generalized) momenta (the so-called {\em canonical variables}) of all physical degrees of freedom to the system's energy and thereby giving the temporal evolution of all the system's degrees of freedom.\footnote{More precisely, they are a system of first-order differential equations expressing the dynamical constraints on the $2n$-dimensional phase space of the system, where $n$ is the number of degrees of freedom.} It turns out that a vast and important class of dynamical physical systems obey Hamilton's equation and can thus be cast as Hamiltonian systems.

GR in its usual formulation is not a Hamiltonian system. At the heart of standard GR, we find the so-called {\em Einstein field equations} which relate the geometry of spacetime, encoded in the metric field, to the distribution of matter and energy in it. They are often interpreted to describe a dynamical interaction between gravity as captured by the metric field and the energy-matter distribution.\footnote{Mathematically, they are a system of ten independent non-linear second-order partial differential equations which reduce to six independent equations when the freedom of choice of spacetime coordinates is taken into account. Four of the ten equations are constraints related to the four-dimensional diffeomorphism invariance, more on which below.} John Wheeler's famous dictum that in GR, mass grips spacetime, telling it how to curve, and spacetime grips mass, telling it how to move,\footnote{Wheeler's quip appears in many of his writings, cf.\ e.g.\ Wheeler (1990, xi).} epitomizes this interpretation of the Einstein field equations as governing the dynamical co-evolution of the spacetime metric and the matter fields. We will have to return to this interpretation of GR in the next section when we discuss the problem of time. 

A solution of the Einstein field equations is a triple $\langle \mathcal{M}, g, T\rangle$ of a four-dimensional differentiable manifold $\mathcal{M}$, a metric field $g$, and a so-called {\em stress-energy tensor} $T$, which expresses mathematically the distribution of matter and energy on the manifold, such that $g$ and $T$ relate to one another in accordance to the Einstein field equations at every point of $\mathcal{M}$. On the face of it, therefore, the Einstein equations are not {\em dynamical} equations; rather, they simply give local conditions on pairs of values of the metric field $g$ and the energy and matter content of the universe as captured by $T$. But formulating GR as a Hamiltonian system requires that it be understood dynamically. In a dynamical theory, one would expect to be able to articulate a well-posed initial value formulation, i.e.\ a formulation of the theory that would enable us to obtain the full dynamical evolution of the physical system for all times given a fully specified set of initial conditions at some time and the dynamical equations. A Hamiltonian formulation of GR affords a natural connection to the initial value problem. This problem, however, is not well-posed in the standard formulation of GR since the thus necessitated split of four-dimensional spacetime into ``space'' that evolves over ``time'' appears to violate the very central lesson of relativity according to which such a split cannot be physically well-motivated. This forced split of spacetime fosters incipient concerns that a dynamical understanding of GR caters to a misinterpretation of it. But even though Hamiltonian GR requires ``foliating'' spacetime into three-dimensional ``spaces'' ordered by a one-dimensional ``time'' parameter, the relativistic lesson of four-dimensionality is mathematically accommodated in the Hamiltonian framework in its constraint equations, on which more below. 

There are various ways in which GR might be ``dynamized'' in order to obtain a Hamiltonian version of the theory. Generally, the idea is to find canonical coordinates that somehow capture a spatial geometry changing over time. To that end, Hamiltonian formulations of GR slice the spacetime into a foliation of three-dimensional spatial hypersurfaces (which are spacelike submanifolds of $\mathcal{M}$). Traditionally, the standard way of doing this is named {\em ADM formalism} after its founders Richard Arnowitt, Stanley Deser, and Charles Misner. The ADM formalism takes the three-metrics induced by $g$ on the spatial hypersurfaces as the ``position'' variables and (a linear combination of components of) the exterior curvature of these hypersurfaces encoding their embedding into the four-dimensional spacetime as ``momentum'' variables, which are canonically conjugate to the three-metrics. Hamilton's equations can then be written down. 

It turns out, however, that they are not, by themselves,  equivalent to Einstein's field equations. For the equivalence to hold, additional equations constraining the relation between the canonical variables must be appended to Hamilton's equations. These constraint equations testify to the fact that initial data cannot be chosen arbitrarily, but must satisfy certain conditions.\footnote{For details on the AMD formalism and how the constraint equations arise there, cf.\ Wald (1984, Chapter 10 and Appendix E.2).} It can be shown that these constraint equations are a mathematical expression of the presence of so-called ``gauge freedom,'' i.e.\ a representational redundancy in the mathematical description of the physical situation.\footnote{Cf.\ W\"uthrich (2006, Section 4.1).} In particular, they arise as a consequence of the fact that the group of four-dimensional diffeomorphisms is the dynamical symmetry group of GR as the principle of general covariance demands.\footnote{A diffeomorphism is a bijective and smooth map between differentiable manifolds whose inverse is also smooth.} General covariance is the requirement that the physics remains unchanged if the fields---including the metric field---are all smoothly pushed around the manifold in the same way. The idea behind the demand for general covariance is thus that although the mathematical expression for the unpushed and the pushed situation will differ, the physical situation is identical in both cases. 

In fact, two (families of) constraint equations arise. The first, encoding the freedom to choose the foliation, is the so-called {\em Hamiltonian constraint}. It turns out that the Hamiltonian of the usual Hamilton's equations is itself a constraint.\footnote{I am glossing over some details here: strictly speaking, it is a linear combination of constraints of both families. But this doesn't change the fact that it is gauge-generating constraint.} Thus, one can see that the absence of an external fiducial time leads to the ``dynamical'' equation being itself a constraint, connected to a freedom of choosing a gauge that has no observable consequences. The second---there are three---, related the freedom to choose spatial coordinates in three-space, are called {\em vector constraints}. This gives a total of four constraint equations.

Once a Hamiltonian formulation of classical GR has been found, i.e.\ once we have identified canonical variables and written down all constraint equations they must satisfy, one can crank the classical theory through the procedure of canonical quantization, as outlined by Paul Dirac (1964). The main idea is to take the canonical variables and turn then into quantum operators acting on a space of quantum states. Their relational structure, as encoded by the Poisson bracket at the classical level, morphs into the canonical commutation relations between the basic operators and the constraint equations become wave equations of constraint operators functionally identical to the classical constraint functions acting on the quantum states. Only those quantum states which satisfy these quantum constraint equations then qualify as physically admissible states. 

Attempts to use the ADM formalism to gain a quantum theory of gravity via canonical quantization have been frustrated by insurmountable technical complications, such as the fact that the constraint equations are non-polynomial. For a moment, then, it looked as if attempts to use canonical quantization to obtain a quantum theory of gravity from GR were fatally doomed. But in the 1980s, new variables were found by Abhay Ashtekar based on work by Amitabha Sen. These {\em Ashtekar variables} simplified the constraint equations significantly,\footnote{Even though the simplification depends on a number of contentious and yet unresolved technical issues.} even though the direct geometric significance of the ADM variables was lost. I will spare you with the mathematical details---these can be found in any decent review of LQG.\footnote{Rovelli (2004) is the standard textbook.} Let me just mention that the basic idea is that the spacetime geometry is captured by a ``triad field'' encoding the local inertial frames defined on the spatial hypersurfaces, rather than the three-metrics. Both approaches equally capture the spacetime geometry and are intertranslatable, even though there is an additional family of constraints in LQG related to internal symmetries. In essence, the move from ADM to Ashtekar variables amounts to a reinterpretation of the Einstein field equations as statements about a ``connection''---a mathematical means of describing what happens to tangent vectors to a manifold that are transported from one point of the manifold to another along a curve---rather than about a metric. The thus reinterpreted general theory of relativity is then subjected to the canonical quantization procedure as outlined above. 

As it turns out, not all constraint equations can easily be solved. In fact, only two of the three families of constraint equations have so far been solved. Let me define the {\em physical Hilbert space} as the space of all quantum states of the theory that solve all the constraints and thus ought to be considered as the {\em physical} states. This implies that the physical Hilbert space of LQG is not yet known. The larger space of states which satisfy the first two families of constraints is often termed the {\em kinematical Hilbert space}. The one constraint that has so far resisted resolution is the Hamiltonian constraint equation with the seemingly simple form $\hat{H} |\psi\rangle = 0$, the so-called {\em Wheeler-DeWitt equation}, where $\hat{H}$ is the Hamiltonian operator usually interpreted to generate the dynamical evolution and $|\psi\rangle$ is a quantum state in the kinematical Hilbert space. Of course, the Hamiltonian operator $\hat{H}$ is a complicated function(al) of the basic operators corresponding to the basic canonical variables. In fact, the very functional form of $\hat{H}$ is  debated as several inequivalent candidates are on the table. Insofar as the physical Hilbert space has thus not yet been constructed, LQG remains incomplete.

\begin{figure}
\centering
\epsfig{figure=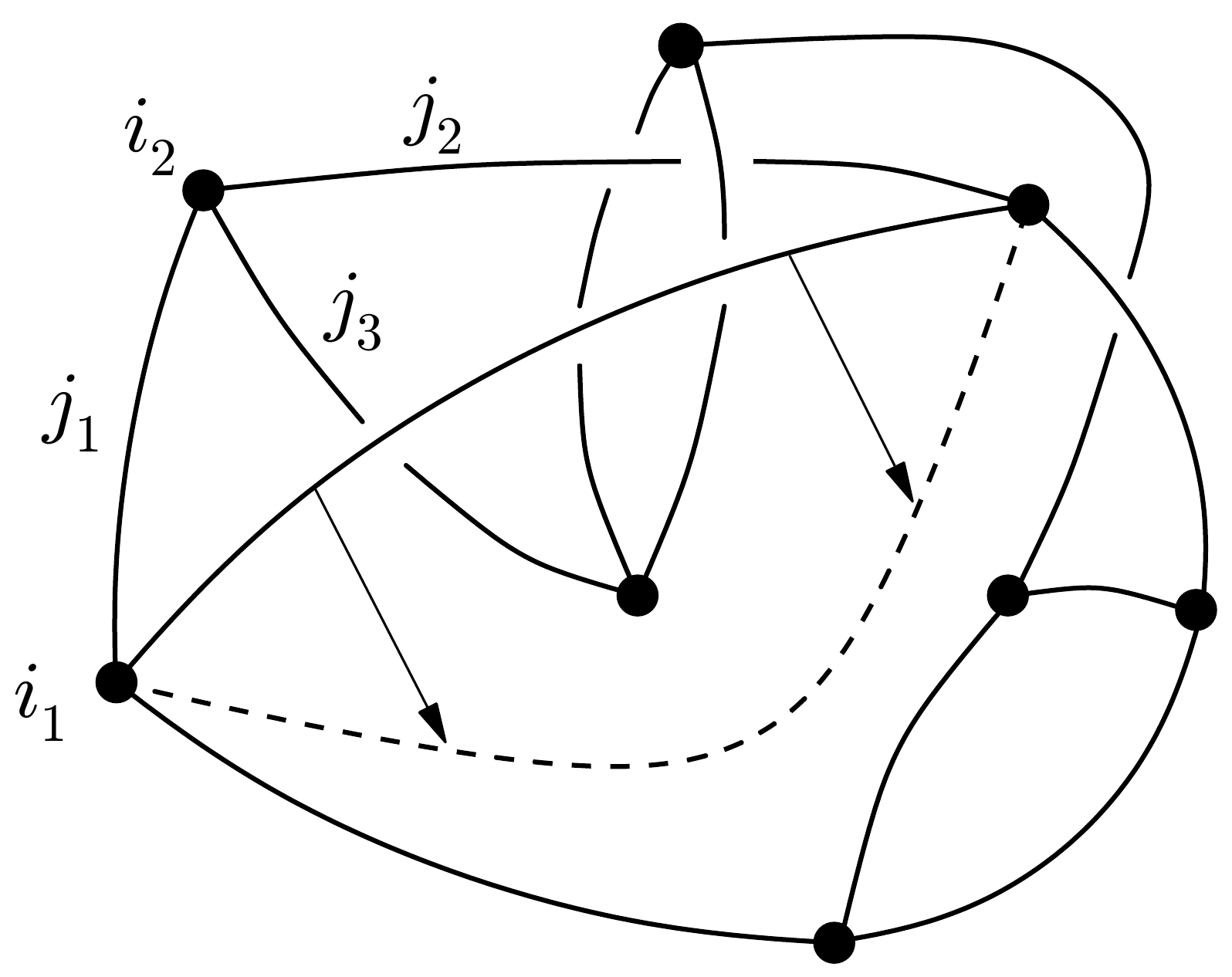,width=0.4\linewidth}
\caption{\label{fig:spinnetwork} Spin network states can be represented by labelled graphs.}
\end{figure}

Since the physical Hilbert space is a subspace of the kinematical Hilbert space, all physical states are also elements in the kinematical Hilbert space. Fortunately, much more is known about this space. Its elements are the {\em spin network states}, the quantum states of the gravitational field, at least as it is spatially distributed. Spin network states can be represented by labelled graphs embedded in some background space, cf.\ Figure \ref{fig:spinnetwork}. Physical space is supposed to be, fundamentally, a spin network state or a quantum superposition of such states.\footnote{More precisely, since spin network states are not invariant under diffeomorphisms, {\em equivalence classes} of spin network states under three-dimensional diffeomorphisms must be taken to encode the fundamental structure of physical space. That spin network states are not diffeomorphism-invariant can be seen from the fact that, strictly speaking, pushing (part of) them around the embedding space without changing their knot structure as indicated by the arrows in Figure \ref{fig:spinnetwork} yields a distinct spin network state each time. But since the invariant knot structure captures the physical situation and not its particular embedding in another space, we must look, again strictly speaking, at what mathematicians call {\em abstract graphs}, i.e.\ equivalence classes of graphs with the same knot structure but embedded differently. This subtle but important point will be ignored in what follows.}

The structure of physical space, therefore, is essentially captured by labelled graphs as in Figure \ref{fig:spinnetwork}. As is indicated there, ``spin''-representations sit on the vertices of the graph (represented by the nodes) as well as on the edges (represented by the lines connecting the nodes). The spin-representations on the vertices, denoted by $i_k$, correspond to quantum numbers indicating the ``size'' of the ``space atoms,'' while those on the edges, labelled by $j_l$, correspond to the ``size'' of the surface connecting adjacent ``chunks'' of space. Spin network states are discrete structures. It can thus be seen that, according to LQG, physical space is granular at the tiny Planck scale. Thus, the smooth space of the classical theory is supplanted by a discrete quantum structure. Hence, space as it figures in our conceptions of the world is an emergent phenomenon, not a fundamental reality. Or so LQG claims.

The two most pressing problems of LQG are our lack of understanding of the dynamics or, equivalently, our inability to solve the Hamiltonian constraint equation as well as our failure to give an account of how the classical smooth spacetime emerges or, equivalently, of how classical gravitational theories such as GR have been as successful as they were. Both of these problems have technical as well as philosophical aspects, and both appear in one guise or another in many of the main approaches to quantum gravity. For instance, the technical problem of solving the Hamiltonian constraint equation in LQG is closely tied to the problem of time, which has many philosophical dimensions. Moreover, to the extent to which string theory contains GR, it must also deal with the problem of time at least at the very general level at which a resolution of the conceptual tension between the pre-relativistic notion of a time external to, and independent of, the physical system at stake employed in quantum theories and in string theory on the one hand and the relativistic reconceptualization of time as a physical actor fused with space and interacting with matter fields and other forms of energy on the other. Naturally, however, the precise form the problem takes will differ, sometimes quite radically, from approach to approach. These two major issues shall be treated separately in the two remaining sections, with an eye on the conceptual and philosophical angles.

\section{The problem of time}
\label{sec:time}

The Presocratic philosopher Parmenides of Elea famously maintained that, fundamentally, the world is an immutable, unchanging, uncreated, indestructible whole. Changes, he argued, are merely apparent and what exists in reality is temporally ``frozen.'' Particularly in recent centuries, few philosophers followed Parmenides in his radical metaphysics. Surprisingly, his brave hypothesis garners support from (Hamiltonian) GR and the quantum theories of gravity based on it. 

Already in (standard) GR, isolating physical time is far from trivial and in general, time---whatever its nature---only induces a preorder of temporal precedence on the set of events on the manifold. A {\em preorder} is a two-place relation $Rxy$ on a set $X$, which is reflexive and transitive, but in general neither weakly antisymmetric nor comparable. Weak antisymmetry is the condition that for any two events $a$ and $b$, if event $a$ temporally precedes $b$ (which is read as to include the possibility of their being simultaneous), and $b$ temporally precedes $a$, then they are simultaneous. Unless time is circular or have some funny topology, weak antisymmetry will hold and is thus generally considered a necessary condition for a legitimately {\em temporal} ordering. But time in GR is not weakly antisymmetric since there may be pairs of events which exemplify the temporal precedence relation in both orders without the events being simultaneous. It is also not comparable because pairs of spacelike related events will not stand in a temporally ordered relation at all, which they would have to in order for comparability to hold. Weak antisymmetry can be salvaged for those spacetimes with topology $\Sigma\times \mathbb{R}$, where $\Sigma$ is any three-dimensional space, $\mathbb{R}$ are the real numbers, and $\times$ designates the Cartesian product. Spacetimes of this topology can thus be split into ``space'' and ``time,'' even though in general there will be infinitely many equally valid ways of performing such a split. Only if one foliation, i.e.\ one particular way of splitting, can be privileged in some physically principled way can comparability, and thus totality, of the ordering relation be regained. 

Since Hamiltonian GR demands that we foliate spacetime into a spatial system which then evolves over time, it can only deal with spacetimes of topology $\Sigma\times\mathbb{R}$. It may thus appear as if the difficulty with time may be {\em alleviated} in Hamiltonian GR (and consequently in approaches to quantum gravity deriving from it) as compared to standard GR. But this impression is deceptive; in fact, things stand much worse. There is a sense in which time completely evaporates, and all physical magnitudes are bound to remain constant over time. 

Concerning the complete disappearance of time in canonical quantum gravity, it was noted by Wheeler and Bryce DeWitt in the 1960s that the basic dynamical equation, their eponymous equation stated above, does not contain a time parameter. Unlike the Schr\"odinger equation
\begin{equation*}
\hat{H}|\psi\rangle = i\hbar \frac{\partial}{\partial t} |\psi\rangle,
\end{equation*}
which gives the dynamics in non-relativistic quantum mechanics, its right-hand side vanishes and there is no time parameter $t$. As far as canonical quantum theories of gravity are concerned, therefore, time simply falls out of the picture. The disappearance of time in the dynamical equations of the canonical approach can be regarded as testimony to the conceptual tension that anyone faces when attempting to marry the stripped-down version of time in general-relativistic spacetimes with the external time, flowing equably and completely independently of the physical systems whose evolution it enables, that we find in quantum mechanics. Perhaps this is a consequence of the fact that time was part of the physical system, viz.\ spacetime, which we quantized. In relativistic physics, there is certainly no external fiducial time with respect to which the dynamics could ``play out.'' But to read off the Wheeler-DeWitt equation that there is no time at all, however, is a bit too quick. It may still be there, of course, but in a way such that the dynamics may not be wearing it on its sleeves, as would only be appropriate in relativistic physics.

Some physicists like Carlo Rovelli and Julian Barbour, however, have embraced the radical conclusion and have consequently attempted to formulate quantum mechanics in a way that does not require an external time clocking down the dynamics, but rather substitutes time by relating events directly to one another.\footnote{For the canonical refence on ``relational quantum mechanics,'' see Rovelli (1996); for a popular account, see Callender (2010).} That there is no time at the quantum level may be acceptable without giving in to the demand that quantum mechanics be relational, as long as GR is found to be capable of describing {\em change} and if we come to understand how classical spacetime emerges from the underlying quantum structure or, equivalently, how classical GR is valid in some low-energy limit as an approximation to LQG. GR can easily deliver on the first part of the condition: while time in general does not even yield an objective, universally valid partial ordering of events, a relational account of change as the variation of properties of physical system along their worldlines can be given in GR. The second part, however, has no easy resolution, as will be seen in Section \ref{sec:emerge}.

But while change can be accounted for in the standard version of GR, there is another aspect of the problem of time which indicates that the appearance one gets in canonical quantum gravity that there is no time is not deceptive after all insofar as there is no change at the most fundamental level of physical reality.\footnote{Cf.\ W\"uthrich (2006, \S 4.3).} In its Hamiltonian formulation, GR can thus not even accommodate change of physical systems in their properties. We see that the problem of time, or at least that of {\em change}, already arises at the classical level, even though only in one formulation of GR. Parmenides, it seems, is vindicated after all: there is no change, everything really is just frozen in---if I may say so---time. 

Formally, this results from the fact that the reparametrization of (space-)time is a gauge symmetry of the theory. Specifically, the dynamical symmetry group of the Einstein equations is Diff$(\mathcal{M})$, the group of four-dimensional diffeomorphisms on $\mathcal{M}$, which gets encoded in the Hamiltonian formulation as constraints that generate these spatiotemporal diffeomorphisms. In other words, change is nothing but a redundancy of the mathematical representation. At heart, the problem results from the demand that all physical magnitudes cannot depend on the mathematical representation---more specifically, the particular coordinate system---employed to describe what is physically really going on. This demand is eminently reasonable, since changes in the representation will have no observable consequences. The physics is there and is the same, regardless of which coordinate system we humans use to describe it. 

Despite this very counterintuitive conclusion, the argument in the preceding paragraph needs to be taken seriously, as John Earman (2002) urges.\footnote{But there are dissenters: cf.\ Maudlin (2002).} Its conclusion has not been reached frivolously and a strong case can be made for each of the steps in the argument. It should thus not be dismissed easily. A straightforward way to brush away the unpalatable conclusion would of course be to reject the Hamiltonian formulation of GR as physically irrelevant, as Tim Maudlin (2002) seems to do. In fact, the consequences into which it forces us may be considered a reductio ad absurdum of the entire approach. But that would be too quick: canonical quantization has been enormously successful in other domains such as in electrodynamics and offers at least a principled and mathematically relatively well-controlled path to the holy grail of quantum gravity. 

There are a number of proposals as to how to deal with the problem of time, as is testified by the overwhelming response of physicists and philosophers alike to the essay competition on the nature of time issued by the Foundational Questions Institute (FQXi).\footnote{Cf.\ e.g.\ Barbour (2008), Kiefer (2008), and Rovelli (2008).} I will not review these proposals here, but suffice it to say that as variegated the responses to the problem of time may be, as strong is the consensus that substantive progress in fundamental physics is unlikely without a sustained reflection on the nature of time and its role in quantum gravity. Whichever position one assumes in the debate, one other thing is also quite clear: that an understanding of how classical spacetime emerges from the fundamental non-spatio-temporal quantum structure will shed light on the problem of time.

\section{The disappearance and re-emergence of spacetime}
\label{sec:emerge}

In string theory as well as in LQG, and in other approaches to quantum gravity, indications are coalescing that not only time, but also space is no longer a fundamental entity, but merely an ``emergent'' phenomenon that arises from the basic physics. In the language of physics, spacetime theories such as GR are ``effective'' theories and spacetime itself is ``emergent,'' much like thermodynamics is an effective theory and temperature is an emergent property at the effective level, as it is built up from the collective behaviour of gas molecules. However, unlike the notion that temperature is emergent, the idea that the universe is not in space and time arguably shocks our very idea of physical existence as profoundly as any scientific revolution ever did. It is not even clear whether we can coherently formulate a physical theory in the absence of space and time.\footnote{Maudlin (2007) doesn't seem to think so.}

Space disappears in LQG insofar as the physical structures it describes bear little, if any, resemblance to the spatial geometries found in GR. As we have seen in Section \ref{sec:lqg}, these structures are discrete and not continuous as classical spacetimes are. They represent the fundamental constitution of our universe that correspond, somehow, to chunks of physical space and thus give rise---in a way yet to be elucidated---to the spatial geometries we find in classical GR. It should be emphasized that the fact that spacetime is replaced by a {\em discrete} structure at the quantum level is a well-established and quite generic consequence of a few basic postulates shared by a rather vast class of quantum theories of gravity, including LQG.\footnote{Cf.\ Smolin (2009, 549).} The conceptual problem of coming to grasp how to do physics in the absence of an underlying spatio-temporal stage on which the physics can play out is closely tied to the technical difficulty of mathematically relating LQG back to GR. Physicists have yet to fully understand how classical spacetimes emerge from the fundamental non-spatio-temporal structure of LQG, and philosophers are only just starting to study its conceptual foundations and the implications of quantum gravity in general and of the disappearance of spacetime in particular.\footnote{As far as I am aware, the philosophical literature on emergence in canonical quantum gravity is exhausted by the two articles by Butterfield and Isham cited in the bibliography and W\"uthrich (2006).} Even though the mathematical heavy-lifting will fall to the physicists, there is a role for philosophers here in exploring and mapping the landscape of conceptual possibilites, bringing to bear the immense philosophical literature in emergence and reduction which offers a variegated conceptual toolbox. Let me say a few preliminary words, in closing, toward mapping a scheme for a resolution of these problems.

To understand how classical spacetime re-emerges from the fundamental quantum structure involves what the physicists call ``taking the classical limit.'' In a sense, relating the spin network states of LQG back to the spacetimes of GR is a reversal of the quantization procedure employed to formulate the quantum theory in the first place. Thus, while the quantization can be though of as the ``context of discovery,'' finding the classical limit that relates the quantum theory of gravity to GR should be considered the ``context of (partial) justification.'' It should be emphasized that understanding how (classical) spacetime re-emerges by retrieving GR as a low-energy limit of a more fundamental theory is not only important to ``save the appearances'' and to accommodate common sense---although it matters in these respects as well---, but must also be considered a methodologically central part of the enterprise of quantum gravity. If it cannot be shown that GR is indeed related to LQG in some mathematically well-understood way as the approximately correct theory when energies are sufficiently low or, equivalently, when scales are sufficiently large, then LQG cannot explain why GR has been empirically as successful as it has been.\footnote{And successful it has been; cf.\ Will (2006).} But a successful theory can only be legitimately supplanted if the successor theory not only makes novel predictions or offers deeper explanations, but is also able to replicate the empirical success of the theory it seeks to replace. 

Ultimately, of course, the full analysis will depend on the full articulation of the theory. But focusing on the kinematical level, and thus avoiding having to fully deal with the problem of time as must Jeremy Butterfield and Chris Isham (1999, 2001), let me apply their concepts to the problem of the emergence of full spacetime, rather than just time as they do. They identify three types of reductive relations between theories: {\em definitional extension}, {\em supervenience}, and {\em emergence}, of which only the last has any chance of working in the case at hand. For Butterfield and Isham, a theory $T_1$ {\em emerges} from another theory $T_2$ just in case there exists either a limiting or an approximating procedure to relate the two theories (or a combination of the two). A {\em limiting procedure} is taking the mathematical limit of some physically relevant parameters, in general in a particular order, of the underlying theory in order to arrive at the emergent theory. A limiting procedure won't work, at least not by itself, due to technical problems concerning the maximal loop density as well as to what essentially amounts to the measurement problem familiar from non-relativistic quantum physics. 

An {\em approximating procedure} designates the process of either neglecting some physical magnitudes, and justifying such neglect, or selecting a proper subset of states in the state space of the approximating theory, and justifying such selection, or both, in order to arrive at a theory whose values of physical quantities remain sufficiently close to those of the theory to be approximated. Note that the ``approximandum,'' the theory to be approximated, in our case will not be GR, but only its vacuum sector of spacetimes of topology $\Sigma \times \mathbb{R}$. One of the central questions will be how the selection of states will be justified. Such a justification would be had if we could identify a mechanism that ``drives the system'' to the right kind of states. Any attempt to finding such a mechanism will foist a host of issues known from the traditional problem of relating quantum to classical mechanics upon us. A candidate mechanism, here and there, is some form of ``decoherence,'' even though that standardly involves an ``environment'' with which the system at stake can interact. But the system of interest in our case is, of course, the universe, which makes it hard to see how there could be any outside environment with which the system could interact. The challenge then is to conceptualize decoherence is a way to circumvents this problem. 

\begin{figure}
\centering
\epsfig{figure=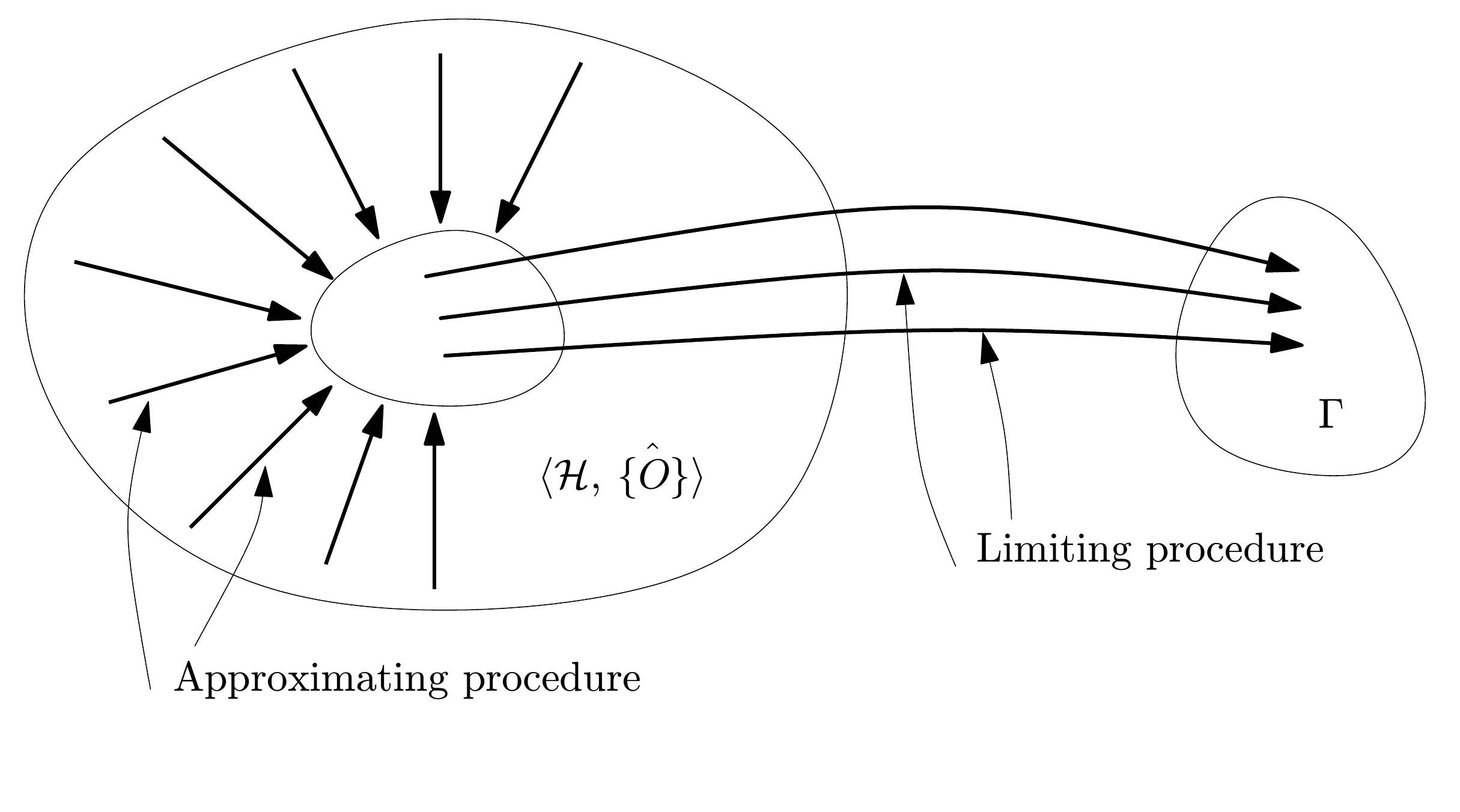,width=0.8\linewidth}
\caption{\label{fig:buttisham} Applying the scheme proposed by Butterfield and Isham (1999).}
\end{figure}

Even though much work remains to be done, both in the technical and the philosophical departments, let me venture the thesis---or should I say ``promissory note''---that at least to the extent to which LQG is a consistent theory, (a close cousin of) GR can be seen to emerge from LQG if a delicately chosen ordered combination of approximations and limiting procedures is applied. The claim is illustrated in Figure \ref{fig:buttisham},
where it can been seen that the idea would be to first apply an approximating procedure at the level of the quantum theory with Hilbert space $\mathcal{H}$ and a set of operators $\{\hat{O}\}$ defined on $\mathcal{H}$ to drive the physical system to a semi-classical subspace which can then be related to the classical space of states $\Gamma$ by a limiting procedure. This is only a very rough sketch, to be sure, and much detail needs to be added, but a beginning is offered in W\"uthrich (2006, Ch.~9). 

Once it is understood how classical space and time disappear in canonical quantum gravity and how they might be seen to re-emerge from the fundamental, non-spatiotemporal structure, the way in which classicality emerges from the quantum theory of gravity does not radically differ from the way it is believed to arise in ordinary quantum mechanics. The project of pursuing such an understanding is of relevance and interest for at least two reasons. First, important foundational questions concerning the interpretation of, and the relation between, theories are addressed, which can lead to conceptual clarification of the foundations of physics. Such conceptual progress may well prove to be the decisive stepping stone to a full quantum theory of gravity. Second, quantum gravity is a fertile ground for any metaphysician as it will inevitably yield implications for specifically philosophical, and particularly metaphysical, issues concerning the nature of space and time. 

\section*{Acknowledgements}

I thank Soazig LeBihan for her kind invitation to contribute to her volume and for her great patience, as well as for her translation. This project has been funded in part by the American Council of Learned Societies through a Collaborative Research Fellowship, the University of California through a UC President's Fellowship in the Humanities, and the University of California, San Diego through an Arts and Humanities Initiative Award.

\end{document}